\documentclass[prb,twocolumn,showpacs,amssymb,amsmath]{revtex4}

\usepackage{graphicx}
\usepackage{bm}
\def\be{\begin{equation}}
\def\ee{\end{equation}}
\def\bea{\begin{eqnarray}}
\def\eea{\end{eqnarray}}
\newcommand{\matr}[4]{{\left(\begin{array}{cc} #1&#2\\#3&#4\\\end{array}\right)}}

\renewcommand{\vec}{\mathbf}
\usepackage{mathtools}

\newcommand{\sgn}{\operatorname{sgn}}

\newcommand{\dd}{\mathrm{d}}

\newcommand{\eps}{\varepsilon}
\newcommand{\emi}{\eps_{\mathrm{min}}}
\newcommand{\esi}{\eps_{\mathrm{size}}}
\newcommand{\echa}{\eps_{\mathrm{charge}}}
\newcommand{\Erf}{\operatorname{Erf}}

\renewcommand{\vr}{\vec{r}}

\newcommand{\vdelta}{\mbox{\boldmath $\delta$}}
\newcommand{\vrho}{\mbox{\boldmath $\rho$}}

\newcommand{\vrhos}{\mbox{\scriptsize \boldmath $\rho$}}
\newcommand{\vsigma}{\mbox{\boldmath $\sigma$}}

\newcommand{\vp}{\vec{p}}

\newcommand{\tr}{\operatorname{tr}}
\newcommand{\Tr}{\operatorname{Tr}}

\begin{document}
\title{Electronic transport in graphene with particle-hole-asymmetric disorder}

\author{Max Hering}
\author{Martin Schneider}
\author{Piet W. Brouwer}
\affiliation{
Dahlem Center for Complex Quantum Systems and Institut f\"ur Theoretische Physik,
Freie Universit\"at Berlin, Arnimallee 14, 14195 Berlin, Germany}
\date{\today}

\begin{abstract}
We study the conductivity of graphene with a smooth but particle-hole-asymmetric disorder potential. Using perturbation theory for the weak-disorder regime and numerical calculations we investigate how the particle-hole asymmetry shifts the position of the minimal conductivity away from the Dirac point $\varepsilon = 0$. We find that the conductivity minimum is shifted in opposite directions for weak and strong disorder. For large disorder strengths the conductivity minimum appears close to the doping level for which electron and hole doped regions (``puddles'') are equal in size.
\end{abstract}

\maketitle

\section{Introduction}
The fascinating electronic transport properties of graphene have been the focus of intense experimental and theoretical investigation over the recent years.\cite{novoselov2004,castroneto2009,geim2009,beenakker2008,peres2010,dassarma2011} A particularly remarkable feature is the appearance of a finite minimum in the electrical conductivity as function of the doping level when the graphene sample is tuned close to the Dirac point where electrons behave as ultra-relativistic particles.\cite{wallace1947} At the Dirac point, the density of states and, hence, the charge-carrier concentration vanish in the absence of disorder. In that case, the minimum conductivity has been shown to originate from evanescent modes in the bulk and its value is found to be $\sigma_0=4 e^2/\pi h$.\cite{ludwig1994,ziegler1998,shon1998,tworzydlo2006,peres2006,katsnelson2006b} 

Disorder that is smooth on the scale of the lattice constant, a condition that is approximately met if the disorder originates from charged impurities at a finite distance from the graphene sheet, increases the value of such minimal conductivity. This at first sight counterintuitive phenomenon can be understood from the observation that a smooth disorder potential always creates a finite carrier concentration, independent of its sign. Therefore, disorder turns a uniform graphene sheet with zero carrier density into a landscape of regions (``puddles'') that are electron or hole doped, separated by a network of carrier-free junctions.\cite{hwang2007,cheianov2007,martin2008,dassarma2011} The effect on the conductivity results from a competition between this increase of carrier density and a simultaneous increase of scattering events among the charged particles. Analytical and numerical calculations have shown that increasing disorder indeed leads to an increase of the conductivity.\cite{bardarson2007,nomura2007,
schuessler2009,dassarma2012}

The existing fully quantum coherent theories of transport near the Dirac point use simplified models for the disorder potential, in which the potential fluctuations are Gaussian. Such Gaussian disorder models are necessarily statistically particle-hole symmetric. The purpose of the present article is to investigate the effect of an inherent statistical asymmetry between electron and hole-like puddles. We note that such asymmetry is relevant for experiments where, {\em e.g.}, the charged impurities in the substrate may have a preferred charge. Besides the experimental relevance, the investigation of the influence of particle-hole-asymmetric disorder on the conductivity of graphene furthermore reveals information about the nature of charge transport in graphene which cannot be inferred from studies with particle-hole-symmetric (Gaussian) disorder. Finally, it has been realized recently, that disorder that does not break certain symmetries on the average may lead to so-called ``statistical topological insulator'' phases, which differ qualitatively from phases in which symmetries are not preserved in a statistical sense.\cite{nomura2008,fulga2014} 

With particle-hole-asymmetric disorder, it becomes interesting to not only look at the {\it magnitude} of the minimal conductivity but also at its {\it position} as a function of the (Fermi) energy. Different models used to qualitatively describe electric transport in graphene predict different values for the doping level at which the minimal conductivity occurs. For example, if electronic transport is determined by percolation physics,\cite{geim2007,cheianov2007} one expects the conductance minimum to appear at a value of the energy $\esi$ for which the total sizes of particle- and hole-like puddles are roughly equal. On the other hand, in a mean-field Boltzmann approach, in which the disorder is characterized by an effective carrier density,\cite{adam2007} the conductivity minimum would be found at an energy $\echa$ where the mean carrier density is zero. Finally, if the electrons are delocalized over distances that are much larger than the correlation length of the disorder, the charge carriers would feel only an effective potential averaged over a large area in space and the conductivity minimum would be expected at the energy $\varepsilon=0$ for which charge carriers feel a potential landscape with zero mean. The same location of the conductivity minimum appears, if one requires that the absolute carrier density is a minimum. While all three values $\esi$, $\echa$ and $\varepsilon=0$ are equal for particle-hole-symmetric potentials, they are in general different if the disorder potential lacks such symmetry. 


The considerations that we present in this article consist of two parts. First, in Sec.\ \ref{AnalSec} we will use an analytical perturbative approach suited to describe the conductivity for finite system sizes and weak disorder strengths. Here we extend an earlier study by Schuessler, Ostrovsky, Gornyi, and Mirlin,\cite{schuessler2009,schuessler2010} that addressed ballistic transport in graphene with Gaussian disorder. Although the perturbation theory is limited to finite system sizes only, we find a size-independent shift of the position conductivity minimum for weak particle-hole-asymmetric disorder. In the second part, we analyze disorder of arbitrary strength using a numerical method. Our main result is that upon introducing particle-hole asymmetry the location of the conductivity minimum shifts in opposite directions for weak and strong disorder. For weak disorder, the numerical results are consistent with the perturbative approach. For strong disorder, we find that the conductivity minimum roughly appears when the Fermi energy is tuned such that electron and hole puddles have the same size. Our numerical findings are discussed in Sec.\ \ref{sec:numerics}. We conclude in Sec.\ \ref{sec:conclusion}. Some technical details are provided in the appendix.

\section{Perturbative approach}
\label{AnalSec}

\begin{figure}[t]
\includegraphics[width=1.5in]{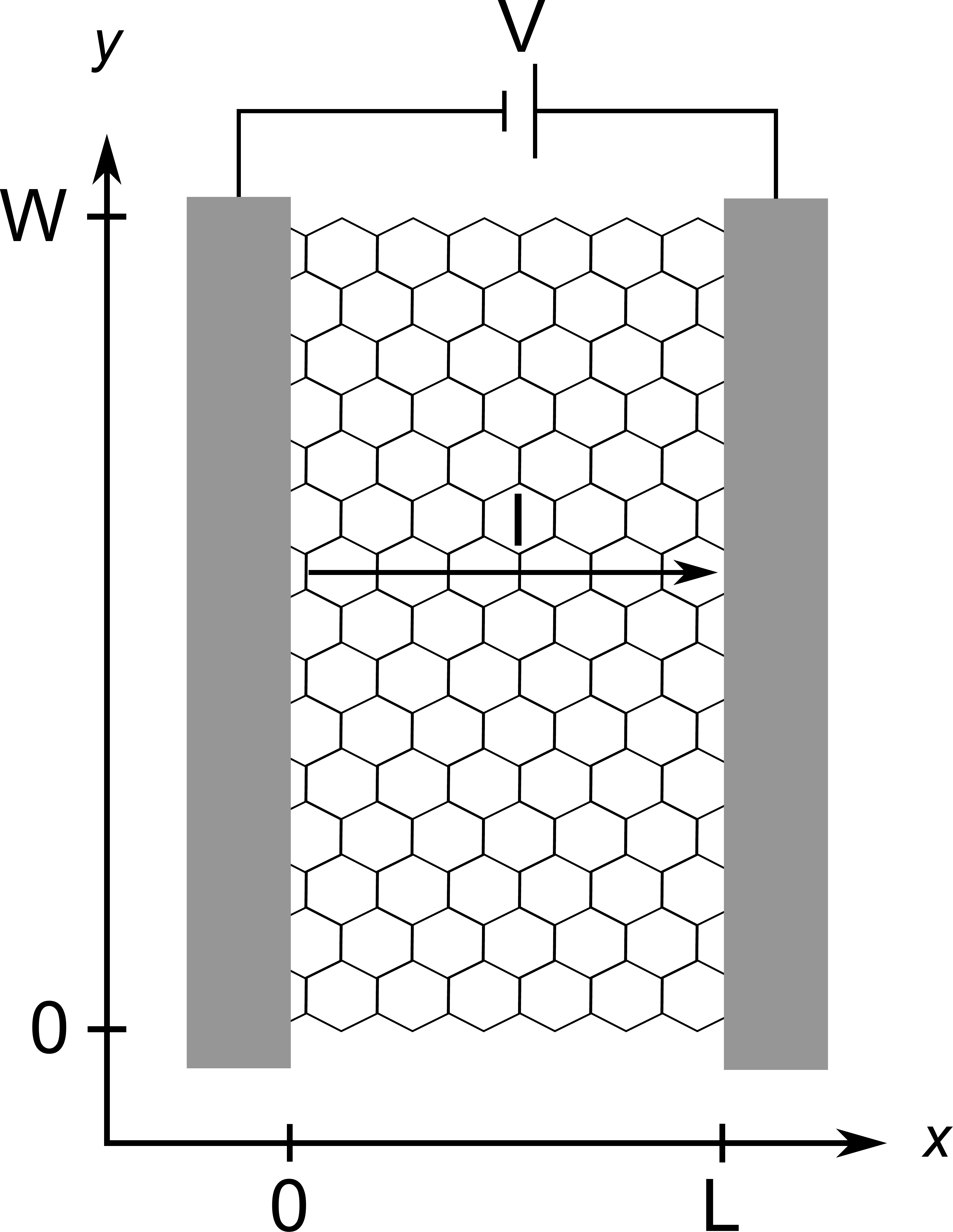}
\caption{Schematic picture of the setup considered for the conductance calculation: A short and wide \((W \gg L)\) graphene strip is connected to two highly doped leads. The leads are modeled by graphene strips with  potential $u \to \infty$. A voltage $V$ is applied to the contacts and we are interested in the resulting current $I = GV$. \label{fig:SetupFig}}
\end{figure}

Our analysis of the weak-disorder regime builds on previous work by Schuessler {\it et al.} \cite{schuessler2010} on the conductivity of graphene in the presence of particle-hole-symmetric disorder. In this section, we first introduce the matrix Green function formalism used in Ref.\ \onlinecite{schuessler2010}, and then apply the formalism to the case of weak particle-hole-asymmetric disorder.

\subsection{Matrix Green Functions and Conductance}

We consider a rectangular sheet of graphene with dimensions $L\times W$ in the short-and-wide limit $W\gg L$, see Fig.\ \ref{fig:SetupFig}. The sample is attached to source and drain contacts at $x=0$ and $x=L$ and exposed to a smooth disorder potential that does not couple the valleys. The Hamiltonian reads
\begin{equation}
 \label{eq:Hamiltonian}
 H=v \vp \cdot \vsigma + V(\vr) - u(x) 
\end{equation}
where $v$ is the velocity of the electrons in graphene, $\vsigma=(\sigma_{x},\sigma_{y})$ is a vector of Pauli-matrices corresponding to the pseudospin degrees of freedom for the two triangular sublattices of graphene, $V(\vr)$ is the disorder potential and $u(x)$ is the potential offset that accounts for the highly doped leads\cite{tworzydlo2006}
\begin{equation}
 u(x)=\begin{cases} 
      0, & 0<x<L \\
      \infty, & \mbox{else}.
      \end{cases}
\end{equation}
To calculate the conductivity of the graphene sample we use the ``Matrix Green function formalism''. This method, originally developed by Nazarov\cite{nazarov1994} and adapted for graphene by Titov {\it et al.} \cite{titov2010}, has been successfully applied in a variety of transport studies for graphene.\cite{schuessler2010,ostrovsky2010,schneider2011,schelter2011,gattenloehner2014} Central object is the matrix Green function (MGF) $\check{G}(\vr,\vr')$, which is defined as
\begin{equation}
\label{eq:matrixgreen}
  \check{G}^{-1} = \matr{\eps-H+i0}{-v_x \delta(x)\sin \frac{\phi}{2}}{-v_x \delta(x-L)\sin \frac{\phi}{2}}{\eps-H-i0},
\end{equation}
where $\eps$ is the electronic energy, $v_x$ is the operator for the $x$-component of the velocity, and $\phi$ is an additional parameter that will be set to zero at the end of the calculation. The parameterization with $\sin \frac{\phi}{2}$ is convenient for the MGF of clean and undoped graphene. For graphene, the Hamiltonian $H$ and the velocity operator $v_x=v\, \sigma_x$ have an additional pseudospin structure such that $\check{G}(\vr,\vr')$ is a $4\times 4$ matrix. We further note that $\eps=0$ corresponds to the case where the Fermi level is tuned to the Dirac point. A finite energy describes a doped graphene sample.

In the Matrix Green function formalism, the conductance $G$ follows from the MGF $\check{G}$ through the relation
\begin{equation}
  \label{eq:condfromomega}
  G=\left. \frac{8 e^2}{h} \frac{\partial^2 \Omega}{\partial \phi^2} \right|_{\phi=0}, \ \ \Omega=\Tr \ln \check{G}
\end{equation}
where the trace $\Tr$ extends over spatial coordinates as well as pseudospin and retarded-advanced space. We have included a factor of 4 for spin and valley degeneracy. The conductivity $\sigma$ is obtained from the conductance $G$ using the relation
\begin{equation}
  \sigma = \lim_{W \gg L} \frac{G L}{W}.
\end{equation}
The limit $W \gg L$ ensures independence of the transverse boundary conditions.

\begin{figure}[t]
\includegraphics[width=2.9in]{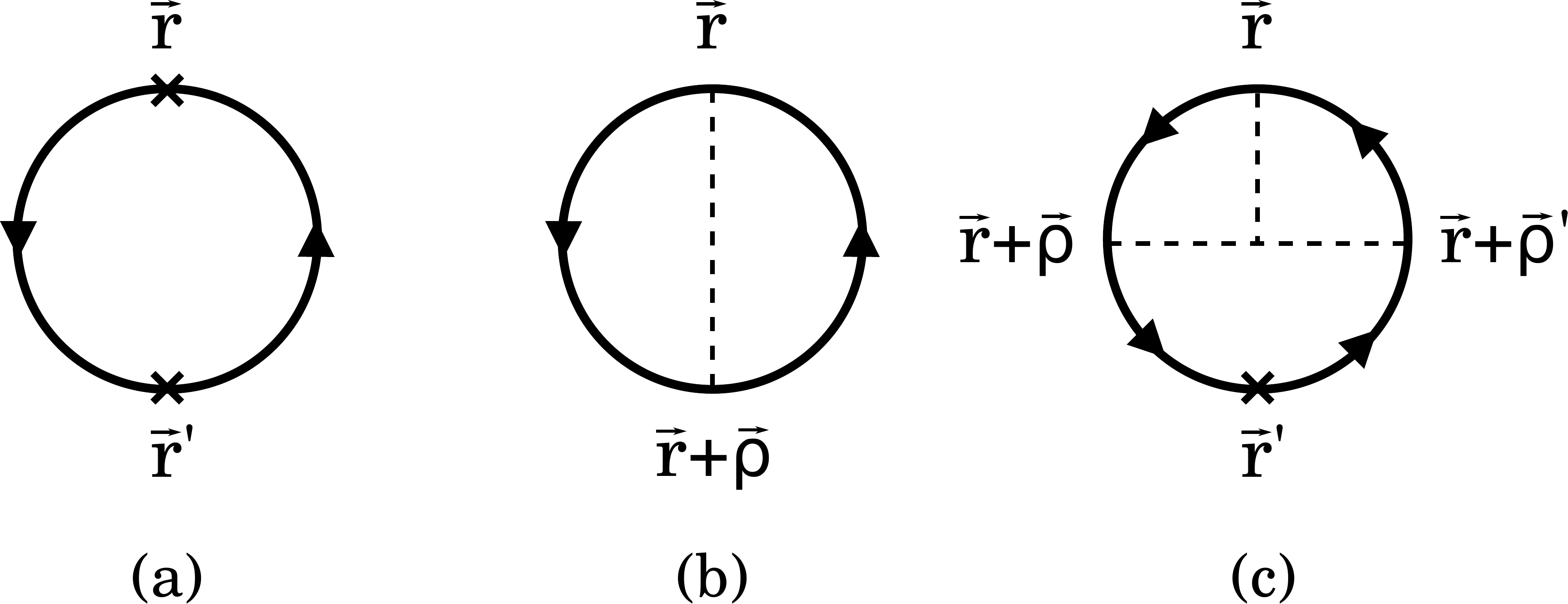}
\caption{Diagrams that contribute to the conductance correction. Crosses represent energy contributions and dashed lines correspond to disorder averages. Diagram (a) yields the second order energy correction to the graphene sample's conductance. Diagram (b) is needed to determine the second order correction in the disorder potential. The two points for which the disorder average is performed are only separated by a small distance of order of the disorder correlation length $\xi$. Finally, (c) generates the conductance correction of a disorder potential's particle-hole asymmetry. Here, three points which are close to each other have to be considered for the third disorder averaged moment. The energy acts at $\vec{r'}$ which does not have to be close to the other coordinates. The resulting conductance contribution is presented in equation (\ref{eq:FinalResult}). \label{SchuesslerDiagrams}}
\end{figure}

Reference \onlinecite{titov2010} gives a closed but lengthy expression for the matrix Green function $\check{G}_0$ in the absence of disorder, from which one obtains the well-known result for the  clean conductance\cite{tworzydlo2006}
\begin{equation}
 G_0=\frac{4e^2}{\pi h}\frac{W}{L}.
\end{equation}
Taking the clean and undoped case as a reference point, the effect of doping and/or disorder can be systematically included using perturbation theory. Hereto, we introduce ${\cal V}=V(\vr)-\eps$ as a perturbation operator that accounts for both doping and disorder effects. Correspondingly we write
\begin{equation}
 \Omega=\Omega_0+\delta\Omega, \quad \Omega_0=\Tr \ln \check{G}_0,
\end{equation}
so that $\Omega_0$ gives the clean-limit conductance $G_0$, while $\delta\Omega$ contains the corrections that exist due to disorder and doping. Explicitly, one has
\begin{equation}\label{eq:OmegaCorrec}
 \delta\Omega=\Tr  \left[ \sum_{n=1}^{\infty} \frac{1}{n}\left({\cal V} \check{G}_0\right)^n \right].
\end{equation}
The latter equation generates diagrams that consist of closed loops of clean MGFs. As shown in Refs.\ \onlinecite{titov2010,schuessler2010}, for such structures, one can work with a simplified version of $\check{G}_0$, 
\begin{align}
\label{eq:CleanGreen}
  \check{G}_0 (x,y;x',0) = \frac{1}{4\hbar v L} \matr{i \cosh \frac{\phi y}{2 L}}{\sinh \frac{\phi y}{2 L}}{\sinh \frac{\phi y}{2 L}}{-i \cosh \frac{\phi y}{2 L}}  \nonumber \\ \times 
  \left[ \frac{1}{\sin \frac{\pi(x+x'+iy \sigma_z)}{2 L}}
  + \frac{1}{\sin \frac{\pi(x-x'+iy \sigma_z)}{2 L}}\sigma_x
  \right].
\end{align}

Before discussing the case of particle-hole-asymmetric disorder, we briefly review some of the results for the case of Gaussian disorder, that were obtained by Schuessler {\it et al.}\cite{schuessler2010} First, we look at the effect of a finite energy $\eps$ inside the sample in absence of any disorder. Here, the first non-vanishing correction arises at second order in $\eps$,
\begin{equation}\label{eq:Omeps}
  \delta \Omega_{\eps}= \frac{\eps^2}{2} \int \dd \vec{r} \int \dd \vec{r'} \tr\left[\check{G}_0(\vec{r},\vec{r'})\check{G}_0(\vec{r'},\vec{r})\right],
\end{equation}
where the trace $\tr$ is performed over the $4\times 4$ matrix structure of the MGF. This correction is represented by the diagram displayed in Fig. \ref{SchuesslerDiagrams}a. The evaluation of the trace and the spatial integrals in Eq.\ \eqref{eq:Omeps} yield the conductance correction\cite{schuessler2010}
\begin{equation}
\delta G_{\eps}=c_1 \left(\frac{\eps L}{\hbar v}\right)^2 G_0
  \label{eq:dGeps}
\end{equation}
with the numerical prefactor $c_1 \approx 0.101\,$. The same result was found from an alternative calculation based on scattering theory.\cite{tworzydlo2006}

We now turn to the effect of a Gaussian disorder potential $V(\vec{r})$ that has a zero mean and is characterized by the correlator
\begin{equation}
 \label{eq:VV}
 \langle  V(\vec{r}) V(\vec{r'}) \rangle = \frac{\alpha (\hbar v)^2}{2\pi \xi^2}e^{-\frac{|\vr- \vr'|^2}{2\xi^2}}
\end{equation}
where $\alpha$ and $\xi$ label the strength and the correlation length of the disorder, respectively. We use angular brackets $\langle \dots \rangle$ to indicate the disorder average. For the description of graphene in terms of a single Dirac cone, it is essential to keep the correlation length $\xi$ finite. To leading order in $\alpha$, one finds for the correction to $\Omega$
\begin{align}
\delta\Omega_{\alpha}&= \frac{1}{2} \int \dd \vec{r} \, \dd \vec{r'} \, \left \langle \tr\left[V(\vec{r})\check{G}_0(\vec{r},\vec{r'})V(\vec{r'})\check{G}_0(\vec{r'},\vec{r})\right] \right \rangle \nonumber \\ &= \frac{\alpha (\hbar v)^2}{4\pi\xi^2} \hspace{-0.5mm} \int \hspace{-0.5mm} \dd \vec{r} \, \dd \vrho \, e^{-\frac{|\vrhos|^2}{2\xi^2}}
  \nonumber \\ & \ \ \ \mbox{} \times
\tr\left[\check{G}_0(\vec{r},\vec{r}+\vrho)\check{G}_0(\vec{r}+\vrho,\vec{r})\right]\label{eq:Omalfa}
\end{align}
see Fig.\ \ref{SchuesslerDiagrams}b for the corresponding diagram. We focus on the limit $\xi \ll L$ for which the correlation length is much shorter than the sample size. In this limit, one can expand the square bracket in the last line of Eq.\ \eqref{eq:Omalfa} for small $\rho$. Concentrating on the term proportional to $\phi^2$, which is relevant for the conductance, one finds that the $\rho$ dependence of the product of the two MGFs in Eq.\ (\ref{eq:Omalfa}) can be neglected. The integrations with respect to $\vrho$ and $\vr$ can thus be performed separately and yield\cite{schuessler2010}
\begin{equation}
\label{EvenCond}
\delta G_{\alpha}=\frac{\alpha}{2\pi} G_0.
\end{equation}
This demonstrates that smooth disorder leads to an increased conductance at the Dirac point. (Reference \onlinecite{schuessler2010} further calculates the correction $\propto \alpha^2$. This result will not be repeated here).

\subsection{Particle-Hole Asymmetric Potentials}

Having reviewed the concept of a matrix Green function and the results for Gaussian disorder, we now apply this method to disorder potentials that have an inherent statistical particle-hole asymmetry. Since a non-vanishing mean of the disorder potential would only correspond to a redefinition of the energy $\varepsilon$, the lowest non-trivial effect of a particle-hole-asymmetric potential arises from its third disorder-averaged moment. To be specific, we assume a correlator of the type
\begin{equation}
 \label{eq:VVV}
  \langle V(\vr) V(\vr') V(\vr'') \rangle  =\beta \frac{3 (\hbar v)^3}{4 \pi^2 \lambda^3} e^{-\frac{|\vr-\vr'|^2+|\vr'-\vr''|^2+|\vr''-\vr|^2}{2\lambda^2}}. 
\end{equation}
where $\beta$ and $\lambda$ quantify the strength and correlation length of the third-order disorder correlator, respectively. Again, we take a finite correlation length into account to avoid any problems that arise at short distances for Dirac-like particles. For the second moment of the disorder potential we continue to use Eq.\ (\ref{eq:VV}).

The effect of such particle-hole-asymmetric potentials can be investigated using perturbation theory following essentially the same steps as for the particle-hole-symmetric potential. First, we note that precisely at the Dirac point ($\eps=0$) there will be no effect linear in $\beta$ since any odd order perturbation in ${\cal V}$ gives a vanishing contribution due to the particle-hole symmetry of the unperturbed problem. In the vicinity of the Dirac point, the correlator (Eq. \eqref{eq:VVV}) will give rise to a term proportional to $\eps \beta$ that leads to a shift of the minimal conductivity away from the Dirac point. Calculating this contribution to $\Omega$ requires the evaluation of the diagram shown in Fig.\ \ref{SchuesslerDiagrams}c,
\begin{multline}
 \label{eq:omegabeta}
 \delta\Omega_{\mathrm{\beta}} = -\eps \int \dd\vr \dd\vr' \dd\vrho \dd\vrho' K(\vr,\vr',\vrho,\vrho') \\
 \times  \langle V(\vr+\vrho) V(\vr) V(\vr+\vrho')\rangle.
\end{multline}
The function $K$ contains the trace over four Green functions
\begin{multline}
  K(\vr,\vr',\vrho,\vrho')
 =\tr \left[\check{G}_0(\vr;\vr+\vrho) \check{G}_0(\vr+\vrho;\vr')\right. \\ \left. \times \check{G}_0(\vr',\vr+\vrho') \check{G}_0(\vr+\vrho';\vr)\right]. 
\end{multline}
We note that a factor $\frac{1}{4}$ in the expansion of Eq.\ \eqref{eq:OmegaCorrec} is canceled by the four possibilities where $\eps$ can be placed for the perturbation ${\cal V}$.  

In the limit $\lambda \ll L$ the main contribution from the disorder correlator comes from values close to $\vrho=\vrho'=0\,$. To perform the integrations, we parameterize $\vrho$ and $\vrho'$ with four-dimensional spherical coordinates,
\begin{align}\label{eq:foursphere}
  \rho_x&=\rho\cos \theta, \nonumber\\
  \rho_y&=\rho\sin \theta \cos \theta_1, \nonumber\\
  \rho'_x&= \rho\sin \theta \sin \theta_1 \cos \varphi,\nonumber\\
  \rho'_y&= \rho\sin \theta \sin \theta_1 \sin \varphi.
\end{align}
To obtain the leading correction of the conductance in the limit $\lambda \ll L$, it is then sufficient to extract the behavior of the function $K$ for small values of $\rho$, where we can focus on the terms proportional to $\phi^2$. Here, it is useful that Eq.\ (\ref{eq:CleanGreen}) for $\check{G}_0$ is written as a tensor product in retarded-advanced and pseudospin space, so that the traces for the two spaces can be carried out separately. We find that
\begin{align}
 \label{eq:Kexpr}
 &\left.  \frac{\partial^2}{\partial \phi^2}K(\vr,\vr',\vrho,\vrho')\right|_{\phi=0} =-\frac{1}{\rho^2}\frac{(y-y')^2}{8\pi^2 (\hbar vL)^4} f(\theta,\theta_1,\varphi)\nonumber \\ 
&\qquad \times\left\lbrace \frac{1}{\cos \frac{\pi }{L} (x-x')-\cosh \frac{\pi }{ L} (y-y')} \right.\nonumber\\
   &\qquad\quad \left. - \frac{1}{\cos \frac{\pi}{L}(x+x')-\cosh \frac{\pi }{L} (y-y')}  \right\rbrace,
\end{align}
up to sub-leading corrections of order $\rho^{-1}$. In Eq.\ \eqref{eq:Kexpr} we abbreviated
\begin{equation}
 f(\theta,\theta_1,\varphi)= \frac{\cos \varphi \cos \theta + \sin \varphi \sin \theta \cos \theta_1}{\sin \theta \sin \theta_1(\cos^2 \theta + \cos^2 \theta_1 \sin^2 \theta)}.
\end{equation}
Although the r.h.s.\ of Eq.\ (\ref{eq:Kexpr}) diverges proportional to $\rho^{-2}$ for small $\rho$, we note that the integration over the four-dimensional sphere contains a factor $\rho^3$ such that the radial integral is well-defined.

We proceed with the calculation of the conductance by combining Eqs.\ \eqref{eq:condfromomega}, \eqref{eq:VVV}, \eqref{eq:omegabeta} and \eqref{eq:Kexpr}. Translational invariance in the transverse direction implies that one integration over a $y$-coordinate simply results in the width $W$ of our sample. The final result reads
\begin{equation}\label{eq:FinalResult}
 \delta G_{\beta}=-c_2  \beta \frac{\eps L^2}{\hbar v \lambda} G_0,
\end{equation}
where the factor $c_2\approx 0.00221$ can be expressed in the form of dimensionless integrals, see Appendix for more details. 

Equation \eqref{eq:FinalResult} is the main result of this Section. As the change \eqref{eq:FinalResult} is linear in energy, together with Eq.\ (\ref{eq:dGeps}) it implies a shift of the position of the minimal conductivity away from $\varepsilon = 0$. Combining the three perturbative corrections to the conductance of graphene in the weak-disorder limit close to the Dirac point, we find
\begin{equation}\label{eq:FinalResultForNum}
 G=\frac{4 e^2}{\pi h} \frac{W}{L} \left [1+\frac{\alpha}{2\pi} - c_2 \beta  \frac{\eps L^2}{\hbar v \lambda} +c_1 \left(\frac{\eps L}{\hbar v}\right)^2 \right ],
\end{equation}
where the numerical constants $c_1$ and $c_2$ are given below Eqs.\ (\ref{eq:dGeps}) and (\ref{eq:FinalResult}), respectively.
{}From this expression, we find that the minimal conductance occurs at the energy
\begin{equation}
  \eps_{\mathrm{min}}= c_{\rm min} \frac{\beta \hbar v}{\lambda},
  \label{eq:epsmin}
\end{equation}
where $c_{\rm min} = c_2/2 c_1 \approx 0.011$.

\section{Numerical Approach}\label{sec:numerics}
We now present a numerical calculation of the conductivity in graphene in the presence of particle-hole-asymmetric random potentials. The numerical simulations can access the regime of weak disorder as well as the regime of strong disorder.

Our calculations are based on a method introduced by Bardarson {\it et al.}\cite{bardarson2007} In this approach, the sample is sliced into many segments of dimension $\delta x \times W$ where the extension $\delta x$ is chosen small enough such that its scattering matrix is determined perturbatively using the Born approximation. After concatenation of the individual scattering matrices, one obtains the transmission matrix $t$ for the entire sample. The conductance $G$ is then obtained from the Landauer formula
\begin{equation}
 G=\frac{4e^2}{h}\tr[t^{\dagger}t].
\end{equation}
For more details on the numerical method we refer to Ref.\ \onlinecite{bardarson2007}. 

To numerically generate a particle-hole-asymmetric disorder potential we start from an auxiliary random function $U(\vr)$ with Gaussian fluctuations, $\langle U(\vr) \rangle = 0$ and
\begin{align}
 \langle U(\vr) U(\vr') \rangle & = \sqrt{\frac{\alpha}{2 \pi}}
  \frac{(\hbar v)^2}{2 \xi^2} e^{-\frac{|\vr-\vr'|^2}{4\xi^2}}.
\end{align}
We then construct the disorder potential $V(\vr)$ as
\begin{equation}
 \label{eq:V}
  V(\vr)=\left[ U(\vr)^2-
  \sqrt{\frac{\alpha}{2 \pi}}\frac{(\hbar v)^2}{2\xi^2}\right]
  \frac{\xi \sqrt{2}}{\hbar v}.
\end{equation}
One easily verifies that $\langle V(\vr) \rangle = 0$, that its second moment obeys Eq.\ (\ref{eq:VV}), and that its third moment is given by Eq.\ (\ref{eq:VVV}), with parameters $\beta= (8/3) \sqrt{2 \pi \alpha^3}$ and $\lambda=\xi \sqrt{2}$. Note that the potential $V(\vr)$ is bounded from below, but not from above, so that it is manifestly particle-hole asymmetric.
\begin{figure}[t]
\includegraphics[width=3.5in]{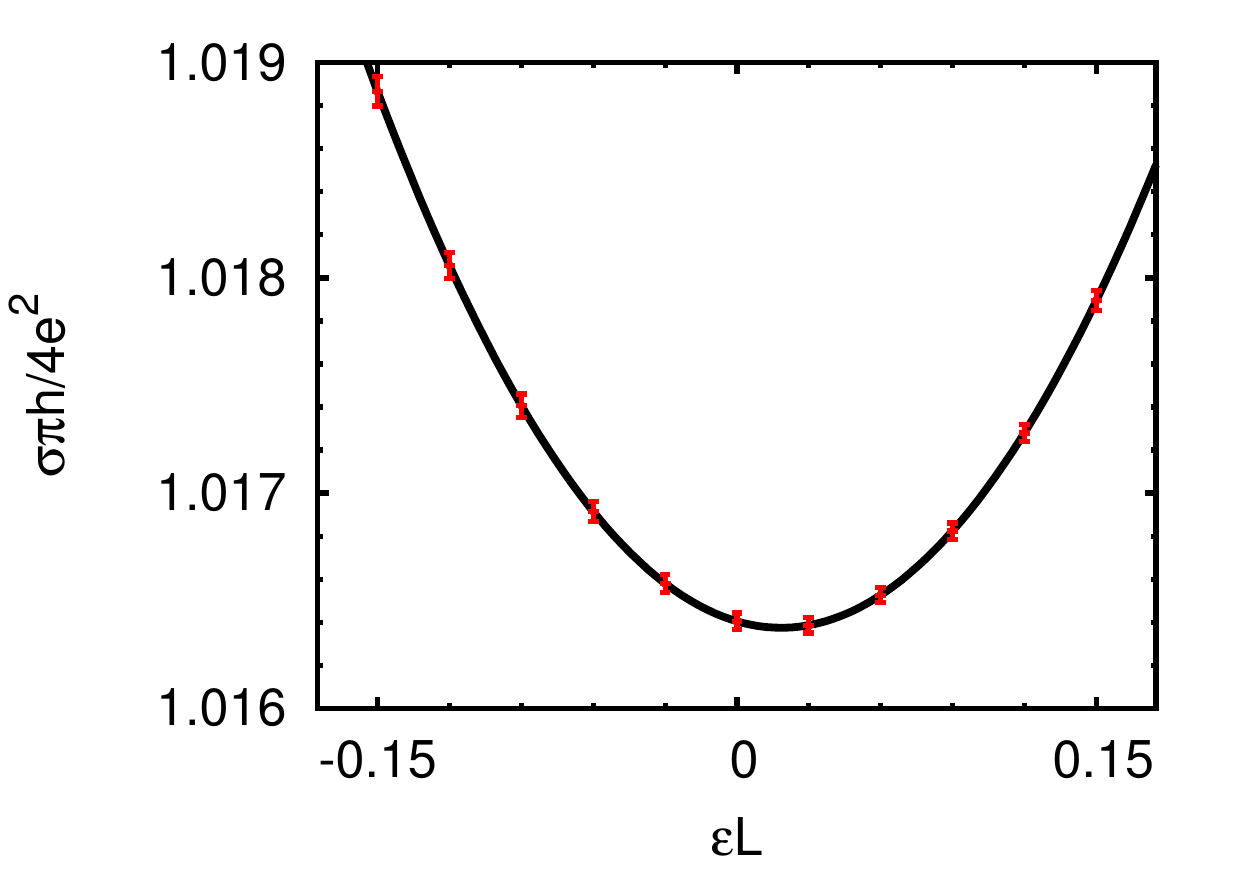}
\caption{(Color online) Parabolic fit of the numerically obtained conductivity data. The parameters are $\alpha=0.16/2\pi$, $L/\xi\sqrt{2}=40$ and $W/L=5$. \label{Fitplot}}
\end{figure}

\begin{figure}[t]
\includegraphics[width=3.5in]{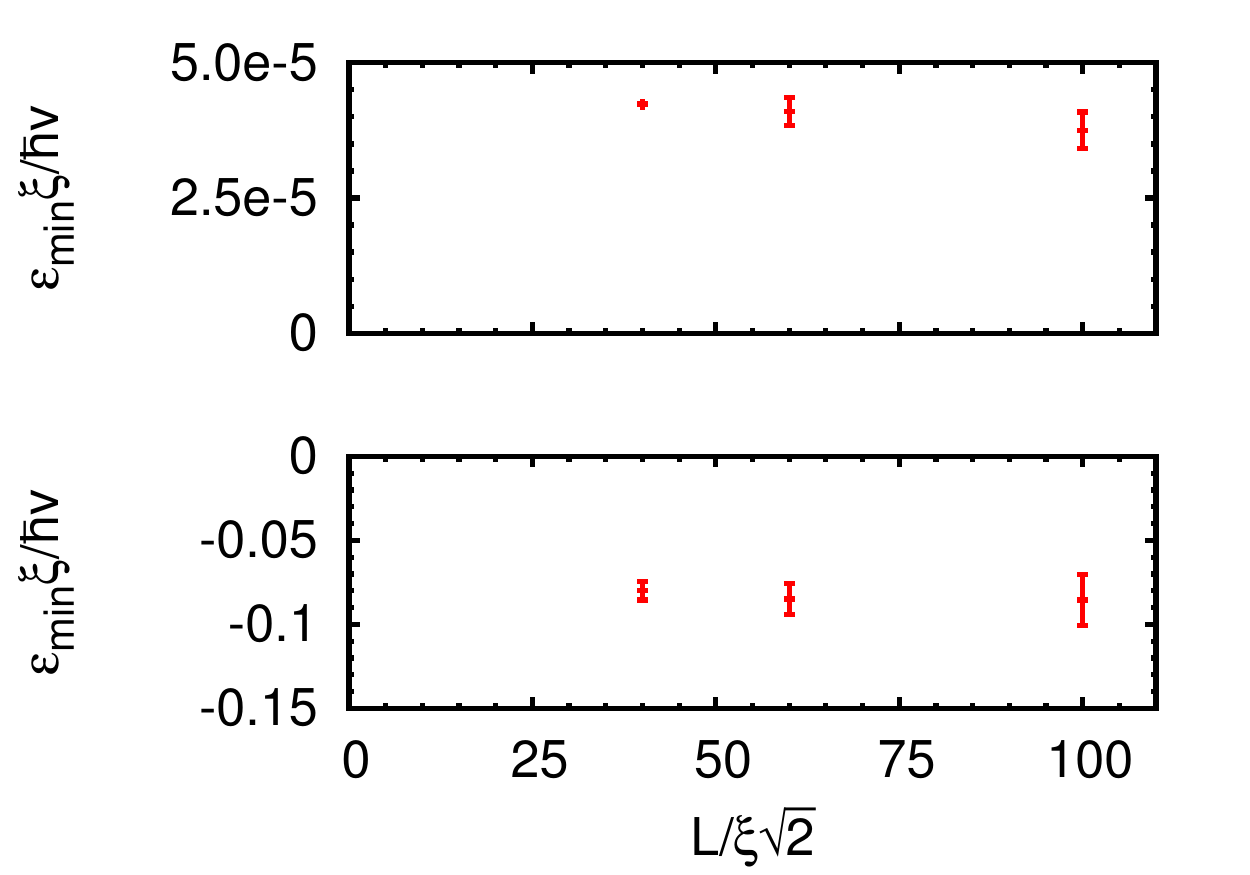}
\caption{(Color online) System-size dependence of the energy value $\emi$ at which the conductivity minimum is found. Upper panel: disorder strength $\alpha=0.04/2\pi$. Lower Panel: $\alpha=16/2\pi$. The aspect ratio $W/L=5$ in both panels. The disorder average was performed for $500$ disorder realizations at $L/\xi\sqrt{2}=40$ and for $200\,\,(25)$ realizations at $L/\xi\sqrt{2}=60\,\,(100)$. \label{SizecheckK0}}
\end{figure}

Using the disorder model (\ref{eq:V}) we numerically calculated the conductance as a function of energy, for various values of the disorder strength $\alpha$. We perform an average over at least $500$ disorder realizations. The aspect ratio $W/L=5$ is chosen large enough to ensure that the conductance is insensitive on the particular boundary conditions in the transverse direction. We then extract the energy $\emi$ at which the minimal conductivity of our graphene sample occurs with a parabolic fit. An example of such a fit is shown in Fig. \ref{Fitplot} for $L/\xi \sqrt{2}=40$ and $\alpha = 0.16/2\pi$. We verified that our result for $\emi$ does not strongly depend on $L$, see Fig. \ref{SizecheckK0} for representative traces for small and large disorder strengths.

\begin{figure}[t]
\includegraphics[width=3.5in]{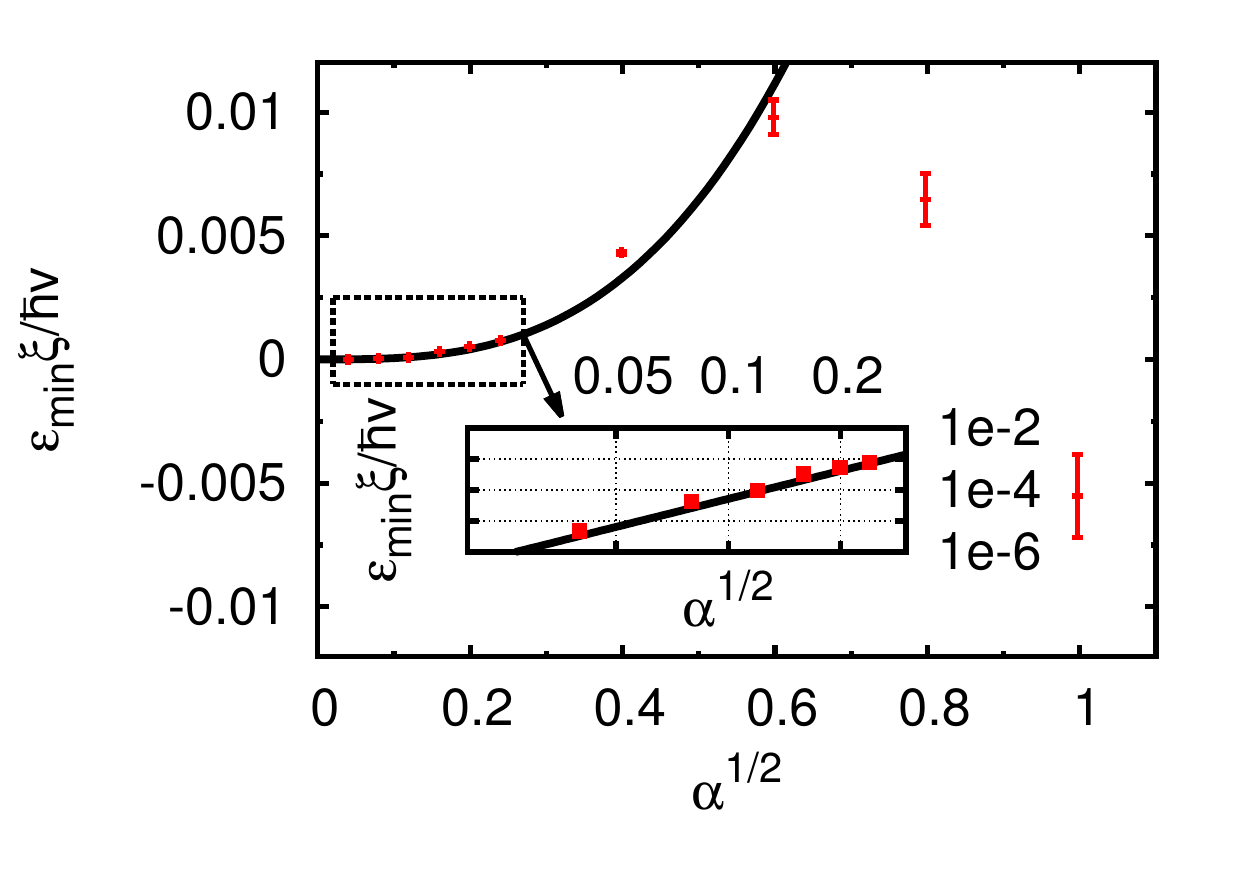}
\caption{(Color online) Location $\emi$ of the conductivity minimum for weak and intermediate particle-hole-asymmetric disorder. The data points are the result of numerical calculations with the disorder potential chosen according to the model (\ref{eq:V}). The result (\ref{eq:epsmin}) of the weak-disorder perturbation theory is indicated with a black line. The inset gives a log-log plot showing the agreement between the numerical results and the perturbation theory. \label{EminSmall}}
\end{figure}

\begin{figure}[t]
\includegraphics[width=3.5in]{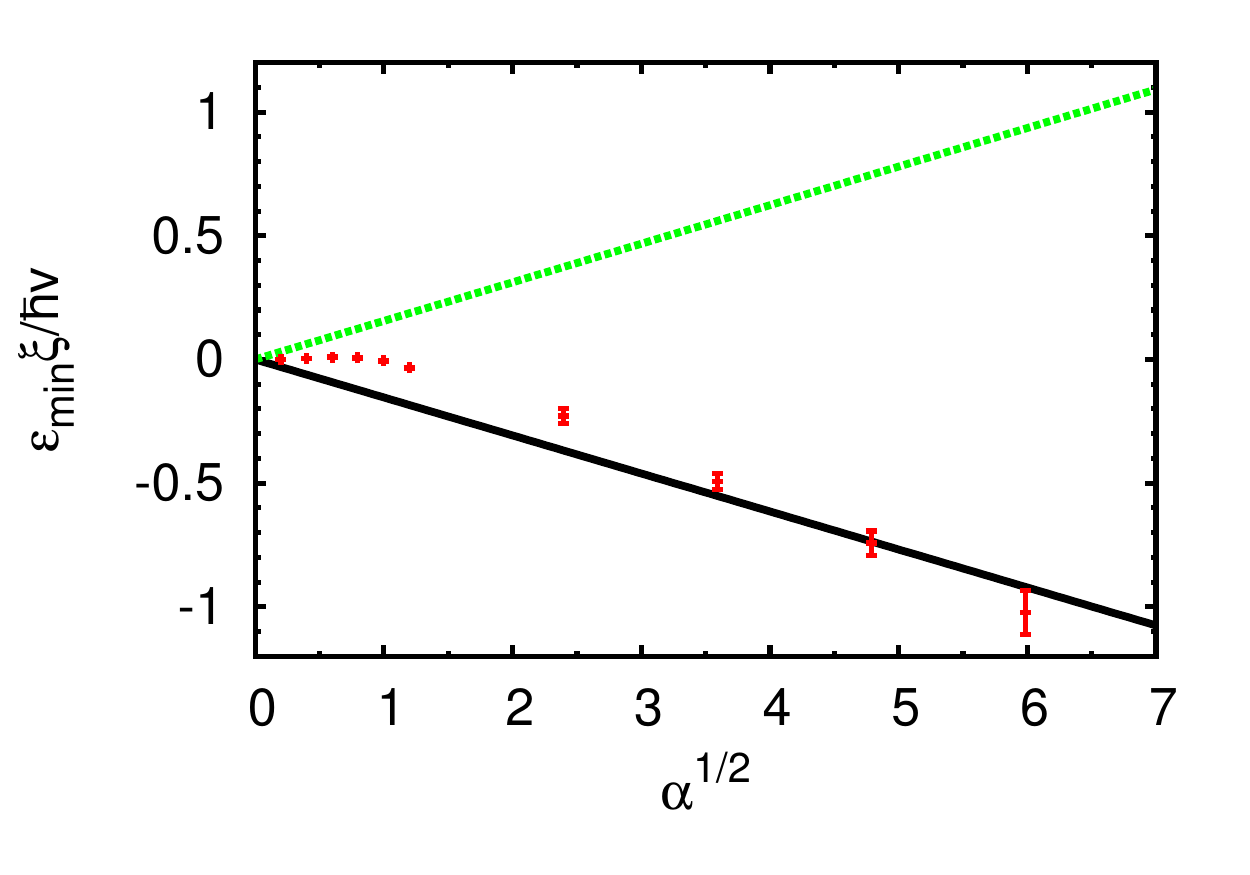}
\caption{(Color online) Location $\emi$ of the conductivity minimum for the full range of disorder strengths. The data points are the results of our numerical calculations, the lines represent the naive estimates $\esi$ (solid, black) and $\echa$ (dashed, green). For large disorder strengths, the numerical data shows a good agreement between $\emi$ and the energy value corresponding to equal sizes of electron and hole puddles inside the sample ($\esi$).\label{EminFullRange}}
\end{figure}

Figure \ref{EminSmall} shows the location of the conductivity minimum for small and intermediate disorder strengths ($\alpha \lesssim 1$). For weak disorder, the numerical results are consistent with the perturbation theory of the previous Section, see the inset of Fig.\ \ref{EminSmall}. Remarkably, beyond the weak disorder regime, the dependence of $\emi$ on the disorder strength is non-monotonous, and $\emi$ becomes negative for $\alpha \gtrsim 1$. This is shown clearly in Fig.\ \ref{EminFullRange}, where we show $\emi$ for the full range of disorder strenghts $\alpha$ considered in the numerical simulations. 

To put these findings in perspective, we consider the three ``naive'' scenarios for the location of $\emi$ mentioned in the introduction: If percolation physics describes the conduction in graphene close to the Dirac point, the conductivity is minimal at the energy $\esi$ for which the sizes of particle and hole-doped regions are equal. If an effective carrier density is the relevant variable that determines electric transport in graphene close to the Dirac point, such minimum would be found at an energy $\eps_{\rm charge}$ for which the total charge of the sample would be zero. Finally, the minimum could also occur for $\eps=0$ for which the landscape $\eps-V(\vr)$ is zero on average. This is the point at which the absolute carrier density has a minimum. It is also the relevant energy if electrons are delocalized over distances larger than the correlation length and would only be affected by an averaged potential. 

We now calculate the energies $\esi$ and $\echa$ for the random potential defined by Eq.\ (\ref{eq:V}). For such calculation we can replace the spatial averages by ensemble averages. As the potential $U(\vr)$ is Gaussian correlated, its probability distribution reads
\begin{equation}\label{Updf}
 \varrho(U)=\frac{2^{1/4}\xi }{\hbar v (\pi \alpha)^{1/4}}
  e^{- (\xi U/\hbar v)^2 (2\pi/\alpha)^{1/2}}.
\end{equation}
The quantity $\esi$ is then determined by the condition that the probabilities that the potential $V(\vr) < \esi$ and that $V(\vr) > \esi$ are precisely equal or, equivalently, that
\begin{equation}
  \int_{V(U) < \esi} dU \varrho(U) = \frac{1}{2}.
\end{equation}
Using Eq.\ (\ref{eq:V}) for $V(U)$ and evaluating the integral gives the condition
\begin{equation}
  \Erf\left( \sqrt{\frac{1}{2} + \sqrt{\frac{\pi}{\alpha}}\frac{ \xi \esi}{\hbar v}}\right ) = \frac{1}{2},
\end{equation}
with the error function $\Erf(x)= (2/\sqrt{\pi}) \int_0^x \dd t\, e^{-t^2}$, so that
\begin{equation}
 \esi   \simeq -0.154 \frac{\hbar v \sqrt{\alpha}}{\xi}.
\end{equation}
In a similar way, one can calculate the energy $\echa$ for which the total charge of the system is zero. For a given value of the disorder potential $V$, the local carrier concentration $n(\eps)$ is given by
\begin{equation}
n(\eps)= \sgn(\eps-V) \frac{\left(\eps-V\right)^2}{\pi(\hbar v)^2}.
\end{equation}
Using this relation, we can calculate $\echa$ from the condition
\begin{align}
  \int_{-\infty}^{\infty} \dd U \,
  n(\echa) \varrho(U) = 0,
\end{align}
where the potential $V$ is given by Eq.\ (\ref{eq:V}).
Solving for $\echa$ gives
\begin{equation}
\echa\simeq  0.156\frac{\hbar v \sqrt{\alpha}}{ \xi}.
\end{equation}
Note that $\echa$ has the same dependence on the disorder strength and correlation length as $\esi$, but that its sign is opposite.
Inspecting Fig.\ \ref{EminFullRange}, we conclude that for large disorder the position of the minimal conductivity agrees reasonably well with the energy scale $\esi$ for which particle and hole puddles equal each other in size. We explain the applicability of the ``percolation-motivated'' picture at large disorder strength with the observation that the charge density in the puddles becomes large for strong disorder, whereas the puddle boundaries remain resistive and are likely to dominate the resistance. For weak disorder none of these three naive limits describes the numerical results well. This is not a large surprise, since transport in graphene in this regime is inherently quantum mechanical and the semiclassical pictures underlying the three naive estimates are likely to fail here.

\section{Conclusion}\label{sec:conclusion}

In this work, we studied how particle-hole asymmetry of the disorder potential affects the conductivity of graphene. Since there are no a priori reasons why a realistic disorder potential should be statistically particle-hole symmetric, it is desirable that simplified models for the disorder do not have a higher symmetry. Moreover, the analysis of the energy at which the conductivity minimum occurs (relative to the Dirac point $\varepsilon=0$) reveals information about the transport mechanisms of graphene close to the Dirac point, which cannot be found in studies with Gaussian disorder, for which the conductivity minimum necessarily occurs at the Dirac point $\varepsilon = 0$.

For a small disorder strength $\alpha$, we calculated $\emi$ from perturbation theory. This result was verified by numerical simulations. Our most remarkable observation is that upon increasing the disorder strength the sign of the energy $\emi$ at which the conductivity minimum occurs changes. At strong disorder, $\alpha \gg 1$, the magnitude and sign of $\emi$ are consistent with a percolation picture, in which the resistivity is dominated by the resistive junctions between otherwise highly conducting electron and hole puddles.

Realistic values for the disorder strength $\alpha$ are in the range $1 \lesssim \alpha \lesssim 4$,\cite{tan2007,adam2008} which, unfortunately, is in the range of intermediate disorder strengths, for which perturbation theory fails, but the strong-$\alpha$ asymptotics has not yet set in.

\acknowledgments
We gratefully acknowledge discussions with Bj\"orn Sbierski, J\"org Behrmann, Christian Fr\"a\ss dorf, and Gregor Pohl. This work is supported by the Alexander von Humboldt Foundation in the framework of the Alexander von Humboldt Professorship, endowed by the Federal Ministry of Education and Research and by the German Research Foundation (DFG) in the framework of the Priority Program 1459 ``Graphene''.

\begin{widetext}
\appendix
\section*{Appendix: Calculation of the numerical constant $c_2$}\label{Append}
In this appendix, we show some details on the calculation of the integrals that lead to the numerical prefactor $c_2$ of Eq.\ \eqref{eq:FinalResult}. As explained in the main text, we transform the integrals over the displacements $\vrho$ and $\vrho'$ to four-dimensional spherical coordinates, see Eq.\ \eqref{eq:foursphere}. The Jacobian for this variable change is
 \begin{equation}
 \int_{-\infty}^{\infty} \dd\vrho \dd\vrho' [...] = \int_0^{\infty} \dd\rho \int_0^{\pi} \dd\theta \int_0^{\pi} \dd\theta_1 \int_0^{2 \pi} \dd\varphi\, \rho^3 \sin^2 \theta \sin \theta_1 [...]\,.
 \end{equation}
Furthermore, the disorder correlator is expressed in terms of the spherical coordinates as
\begin{equation}
 \langle V(\vr+\vdelta) V(\vr) V(\vr+\vdelta')\rangle 
=\beta \frac{3 (\hbar v)^3}{4 \pi^2 \lambda^3} e^{-\rho^2\left[1-\frac{1}{2}\cos\varphi\sin 2\theta \sin \theta_1-\frac{1}{2}\sin\varphi\sin^2\theta \sin 2\theta_1\right]/\lambda^2}\,.
\end{equation}
In order to obtain Eq.\ \eqref{eq:FinalResult}, one utilizes Eqs.\ \eqref{eq:condfromomega}, \eqref{eq:VVV}, \eqref{eq:omegabeta} and \eqref{eq:Kexpr} from the main text. Translational invariance in the $y$-direction implies that one $y$-integral results in the system width $W$. We rescale the remaining $y$-integration as well as the two $x$-integrations with $L$ and the $\rho$-integration with $\lambda$ to obtain Eq.\ \eqref{eq:FinalResult} from the main text. The numerical constant is then given by
\begin{equation}
 c_2={\cal I}_{\vrhos} {\cal I}_{\vr}
\end{equation}
where
\begin{eqnarray}
 {\cal I}_{\vrhos} &=& \int_0^{\infty} \dd \rho \int_0^{\pi} \dd \theta \int_0^{\pi} \dd \theta_1 \int_0^{2 \pi} \dd \varphi\, \rho \sin\theta   \frac{\cos \varphi \cos \theta + \sin \varphi \sin \theta \cos \theta_1}{\cos^2 \theta + \cos^2 \theta_1 \sin^2 \theta} e^{-\rho^2\left[1-\frac{1}{2}\cos\varphi\sin 2\theta \sin \theta_1-\frac{1}{2}\sin\varphi\sin^2\theta \sin 2\theta_1\right]}, \nonumber 
\\
 {\cal I}_{\vr} &=&\frac{3}{16\pi^3}\int_{-\infty}^{\infty} \dd y \int_0^1 \dd x \dd x' \left\lbrace  \frac{y^2}{\cosh \pi y-\cos \pi (x-x')} - \frac{y^2}{\cosh \pi y-\cos \pi(x+x')}\right\rbrace.
\end{eqnarray}

We first discuss the calculation of $\cal{I}_{\vrhos}$. The integration over the variable $\rho$ is straightforward and gives
\begin{equation}
 {\cal I}_{\vrhos}= \int_0^{\pi} \dd\theta \int_0^{\pi} \dd \theta_1 \int_0^{2 \pi} \dd \varphi\,  \sin\theta  \frac{\cos \varphi \cos \theta + \sin \varphi \sin \theta \cos \theta_1}{(2-\cos\varphi\sin 2\theta \sin \theta_1-\sin\varphi\sin^2\theta \sin 2\theta_1)(\cos^2 \theta + \cos^2 \theta_1 \sin^2 \theta)}.
\end{equation}
Proceeding with the $\varphi$-integration, one writes
\begin{equation}
 \frac{\cos \varphi \cos \theta + \sin \varphi \sin \theta \cos \theta_1}{\cos^2 \theta + \cos^2 \theta_1 \sin^2 \theta}= \mathrm{Re}\left[ \frac{e^{i\varphi}}{\cos \theta + i \sin \theta \cos \theta_1}\right].
\end{equation}
Using the identity
\begin{equation}
 \int_0^{2\pi} \dd\varphi \, \frac{e^{i\varphi}}{1-\alpha \cos\varphi - \beta \sin\varphi}=2\pi(\alpha+i\beta)\frac{1-\sqrt{1-\alpha^2-\beta^2}}{(\alpha^2+\beta^2)\sqrt{1-\alpha^2-\beta^2}},
\end{equation}
for $|\alpha|,|\beta|<1/2$,
we then find that
\begin{equation}
 {\cal I}_{\vrhos}=\pi \int_0^{\pi} \dd\theta \int_0^{\pi} \dd\theta_1 \pi\sin^2\theta \sin \theta_1 \frac{1-\sqrt{1-g(\theta,\theta_1)}}{g(\theta,\theta_1)\sqrt{1-g(\theta,\theta_1)}},
\end{equation}
where we abbreviated
\begin{equation}
  g(\theta,\theta_1)=\sin^2\theta \sin^2\theta_1\left(\cos^2\theta + \sin^2\theta \cos^2\theta_1\right).
\end{equation}
This integrand is now a well-behaved function of $\theta$ and $\theta_1$ and the remaining integrals can be performed numerically, with the result
\begin{equation}
 {\cal I}_{\vrhos}\approx 5.67862.
\end{equation}

For the calculation of $\cal{I}_{\vr}$ we start with the $y$-integration, where we use the identity
\begin{equation}
 \int_{-\infty}^{\infty} \dd y \frac{y^2}{\cosh y-\cos \Phi}=\frac{2}{3} \frac{\Phi(\Phi-\pi)(\Phi-2\pi)}{\sin \Phi},
\end{equation}
which is valid for $0\leq\Phi\leq 2\pi$, and can be continued to the entire real axis via the periodicity in $\Phi$. 
This leaves us with the integral
\begin{equation}
 {\cal I}_{\vr}=\frac{1}{8\pi^3} \int_0^1 \dd x \dd x' \left\lbrace  \frac{|x-x'|(|x-x'|-1)(|x-x'|-2)}{\sin \pi |x-x'|} -\frac{(x+x')(x+x'-1)(x+x'-2)}{\sin \pi(x+x')}\right\rbrace.
\end{equation}
For the remaining integrations over $x$ and $x'$, it is customary to shift to the sum and the difference of these variables. Since each term depends only on one of these quantities, the integration over the other variable is straightforward. After some algebra, one obtains

\begin{eqnarray}
  {\cal I}_{\vr} &=&
  \frac{1}{2 \pi^3} \int_0^1 \dd u \frac{u(1-u)(u-\frac{1}{2})^2}{\sin \pi u}
  \nonumber \\ &=&
  \frac{35 \zeta(3)}{8 \pi^6} - \frac{93 \zeta(5)}{2 \pi^8}
  \nonumber \\ &\approx& 0.000388585,
\end{eqnarray}
where $\zeta$ is the Riemann-$\zeta$ function.

\end{widetext}


\end{document}